\documentclass[sigconf]{acmart}

\AtBeginDocument{%
  }

\copyrightyear{2026}
\acmYear{2026}
\setcopyright{cc}
\setcctype{by}
\acmConference[KDD 2026]{Proceedings of the 32nd ACM SIGKDD Conference on Knowledge Discovery and Data Mining V.2}{August 9--13, 2026}{Jeju Island, Republic of Korea.}
\acmBooktitle{Proceedings of the 32nd ACM SIGKDD Conference on Knowledge Discovery and Data Mining V.2 (KDD 2026), August 9--13, 2026, Jeju Island, Republic of Korea}
\acmISBN{979-8-4007-2259-2/2026/08}
\acmDOI{10.1145/3770855.3818457}

\usepackage{multirow}
\usepackage{enumitem}
\usepackage{marvosym}

\makeatletter
\let\acm@orig@fnsymbol\@fnsymbol
\renewcommand{\@fnsymbol}[1]{\ifnum#1=3 \Letter\else\acm@orig@fnsymbol{#1}\fi}
\makeatother

\begin{document}

\title{OneRank: Unified Transformer-Native Ranking Architecture for Multi-Task Recommendation}
\author{Jiakai Tang}
\authornote{Equal contribution.}
\authornote{Also affiliated with Beijing Key Laboratory of Research on Large Models and Intelligent Governance, and Engineering Research Center of Next-Generation Intelligent Search and Recommendation, MOE.}
\affiliation{%
  \institution{Gaoling School of Artificial Intelligence, Renmin University of China}
  \city{Beijing}
  \country{China}
}
\email{tangjiakai5704@ruc.edu.cn}

\author{Sunhao Dai}
\authornotemark[1]
\authornotemark[2]
\affiliation{%
  \institution{Gaoling School of Artificial Intelligence, Renmin University of China}
  \city{Beijing}
  \country{China}
}
\email{sunhaodai@ruc.edu.cn}

\author{Kun Wang}
\affiliation{%
  \institution{Shopee Pte. Ltd.}
  \city{Beijing}
  \country{China}
}
\email{wk1135256721@gmail.com}

\author{Zhiluohan Guo}
\affiliation{%
  \institution{Shopee Pte. Ltd.}
  \city{Shanghai}
  \country{China}
}
\email{guozhiluohan@gmail.com}

\author{Yu Zhao}
\affiliation{%
  \institution{Shopee Pte. Ltd.}
  \city{Beijing}
  \country{China}
}
\email{zy18600749420@gmail.com}

\author{Cong Fu}
\affiliation{%
  \institution{Shopee Pte. Ltd.}
  \city{Singapore}
  \country{Singapore}
}
\email{fc731097343@gmail.com}

\author{Kangle Wu}
\affiliation{%
  \institution{Shopee Pte. Ltd.}
  \city{Singapore}
  \country{Singapore}
}
\email{kangle.wu@shopee.com}

\author{Yabo Ni}
\affiliation{%
  \institution{Nanyang Technological University}
  \city{Singapore}
  \country{Singapore}
}
\email{yabo001@e.ntu.edu.sg}

\author{Anxiang Zeng}
\affiliation{%
  \institution{Nanyang Technological University}
  \city{Singapore}
  \country{Singapore}
}
\email{zeng0118@ntu.edu.sg}

\author{Xu Chen}
\authornote{Corresponding authors.}
\authornotemark[2]
\affiliation{%
  \institution{Gaoling School of Artificial Intelligence, Renmin University of China}
  \city{Beijing}
  \country{China}
}
\email{xu.chen@ruc.edu.cn}

\author{Jun Xu}
\authornotemark[3]
\authornotemark[2]
\affiliation{%
  \institution{Gaoling School of Artificial Intelligence, Renmin University of China}
  \city{Beijing}
  \country{China}
}
\email{junxu@ruc.edu.cn}

\renewcommand{\shortauthors}{Jiakai Tang et al.}
\renewcommand{\authors}{Jiakai Tang, Sunhao Dai, Kun Wang, Zhiluohan Guo, Yu Zhao, Cong Fu, Kangle Wu, Yabo Ni, Anxiang Zeng, Xu Chen, Jun Xu}

\begin{abstract}

Multi-task learning (MTL) is essential in recommender systems to enable complementary learning among diverse user feedback. While modern industrial practices have shifted from DNNs to Transformer-centric architectures to strengthen sequence modeling and scaling capacity, they still decouple feature encoding from multi-task prediction, treating the Transformer as a task-agnostic encoder. This design fundamentally limits the performance and scalability by (1) creating an information bottleneck under heterogeneous task objectives, (2) inducing gradient interference that leads to the seesaw phenomenon, and (3) forcing a dataflow transition in which attention-based, context-adaptive representation learning is converted to static feed-forward task prediction with incompatible information read–write dynamics.

In this paper, we propose OneRank, a Transformer-native multi-task ranking framework that eliminates the encoder–predictor separation and introduces task-private channels for both forward representation learning and backward optimization, enabling task-specialized learning while minimizing inter-task interferences.
In the forward pass, OneRank learns task-specific representations in a bottom-up manner through task-conditioned information selection, candidate–aware contextualization, and controlled cross-task interaction. In the backward pass, cross-task gradient detachment isolates task-private parameter updates from shared knowledge extraction modules, preventing negative transfer. Finally, we replace static task-specific MLP scorers with a dynamic matching-based scoring formulation for context-aware personalized ranking.
By internalizing multi-task reasoning pathways within the Transformer stack, OneRank establishes a new architectural paradigm with unified modeling and scalable computation design.
Extensive offline and online experiments on large-scale industrial datasets demonstrate that OneRank significantly outperforms state-of-the-art baselines across multiple tasks, with substantial improvements in ranking effectiveness while maintaining computational efficiency.

\end{abstract}



\maketitle

\section{Introduction}
\label{sec:intro}

Multi-task learning (MTL) has become the de facto paradigm in modern recommender systems~\cite{wang2023multi,zhang2025advances,ning2010multi}, where joint modeling of dense-but-noisy and sparse-yet-informative feedback capture complementary aspects of user preferences. Earlier Deep Learning Recommendation Models (DLRM)~\cite{ma2018entire,wen2020entire,zhang2020large,wu2022multi} primarily exploit task dependencies in two ways: (i) \emph{explicit dependency modeling} through structured knowledge transfer, such as ESMM~\cite{ma2018entire}, ESCM~\cite{wang2022escm2}, AITM~\cite{xi2021modeling}, and ResFlow~\cite{fu2024residual}; and (ii) \emph{implicit knowledge sharing} via dynamic routing and expert balancing mechanisms, as exemplified by MMoE~\cite{ma2018modeling} and PLE~\cite{tang2020progressive}. Motivated by advances in large language models, recent work~\cite{zhu2025rankmixer,chai2025longer,zhang2025onetrans,han2025mtgr,xu2025climber, dai2025onepiece} has shifted toward Transformer-centric architectures to exploit their strong sequence modeling capability and favorable scaling behavior.

However, this transition does not constitute a fundamental architectural shift. Existing approaches largely retain an encoder–predictor design, which can be formalized as $\mathcal{G}(\mathbf{Z}=\mathcal{F}(\mathbf{X}))$, where $\mathcal{F}(\cdot)$ maps raw inputs $\mathbf{X}$ to a shared, task-agnostic representation $\mathbf{Z}$, and $\mathcal{G}(\cdot)$ denotes task-specific predictors operating on $\mathbf{Z}$. This paradigm has three fundamental limitations as follows:

$\bullet$ \textbf{First}, the shared representation $\mathbf{Z}=\mathcal{F}(\mathbf{X})$ creates a task-agnostic \emph{information bottleneck}, in which task-specific signals are entangled with shared knowledge and lose their identity. Replacing $\mathcal{F}(\cdot)$ with a Transformer increases encoding capacity but does not change this structural constraint, leaving downstream predictors $\mathcal{G}(\cdot)$ to disentangle task-specific information from a fused embedding.
This architectural choice severely limits the model's ability to learn task-specific representations early in the pipeline, forcing complex disentanglement to occur at the prediction stage where modeling capacity is typically constrained.

$\bullet$ \textbf{Second}, shared-bottom architectures are prone to the \emph{seesaw phenomenon}~\cite{he2022metabalance}, where conflicting gradients on shared parameters may improve one task while degrade others. This occurs because task-agnostic bottleneck $\mathbf{Z}$ lack explicit mechanisms to separate task-specific optimization directions backpropagating through $\mathcal{F}$.

$\bullet$ \textbf{Third}, the encoder–predictor separation forces a fundamental dataflow and design pattern transition, in which context-adaptive learning in $\mathcal{F}(\cdot)$ is handed off to static feed-forward task predictors in $\mathcal{G}(\cdot)$. Specifically, Transformers perform iterative, context-dependent information routing through attention, while DNN-based predictors seek a static, global non-linear decision boundary with limited ability to adapt to dynamic user context. This mismatch in design paradigms disrupts end-to-end task reasoning and coherent computation scaling.

\begin{figure}
    \centering
    \includegraphics[width=\linewidth]{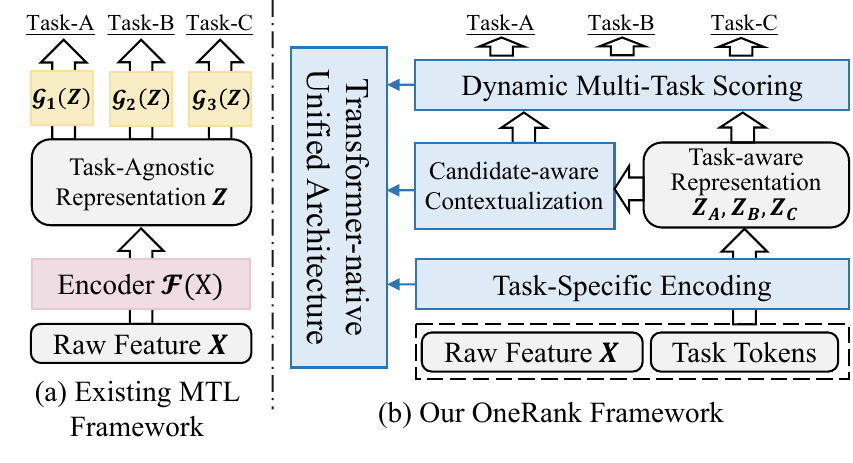}
    \vspace{-20pt}
    \caption{Architectural comparison between (a) traditional encoder–predictor paradigm and (b) our proposed OneRank framework. OneRank internalizes multi-task reasoning within the Transformer-native stack, enabling task-specialized representation learning, dynamic context-aware ranking, and controlled cross-task knowledge transfer without architectural transitions.}
    \vspace{-15pt}
\end{figure}

To address these limitations, we propose \textbf{OneRank}, a Transformer-native multi-task ranking framework that removes the encoder--predictor split by internalizing multi-task reasoning within the Transformer stack itself.
\textbf{In the forward pass}, OneRank builds task-private channels alongside task-shared pathways in a bottom-up manner: at the input level, task-specific tokens with mutual invisibility enable early specialization; at the intermediate level, candidate-aware contextualization aggregates cross-candidate signals via situational descriptors; at the prediction level, controlled cross-task relational attention selectively injects domain-specific task dependencies when beneficial. 
\textbf{In the backward pass}, OneRank employs strategic gradient detachment to block cross-task gradient flow through attention, isolating task-specific parameter updates from shared components and effectively \emph{turning cross-task attention into a read-only memory for knowledge transfer}. 
\textbf{At prediction time}, OneRank replaces static global MLP scorers with a dynamic matching-based formulation, where task-aware global representations are directly matched against context-conditioned candidate embeddings through inner product similarity. 
This unified design enables context-aware and task-adaptive ranking without introducing extra architectural components.

In summary, our contributions are as follows:
\begin{itemize}[leftmargin=*,topsep=2pt]
    \item We identify critical limitations in Transformer-based MTL recommenders and propose OneRank, a unified framework internalizing multi-task reasoning within a Transformer-native design.
    \item We design a bottom-up, task-aware computation paradigm that supports task specialization, task-wise global representation with contextualization, controlled cross-task interaction, and stable optimization, mitigating the information bottleneck and inter-task gradient interference.
    \item We replace static MLP-based prediction heads with a Transformer-native matching formulation, enabling context-aware and task-adaptive ranking within a consistent representation space.
    \item Extensive offline and online A/B testing experiments on large-scale industrial datasets show significant improvements in both effectiveness and efficiency over state-of-the-art baselines.
\end{itemize}

\section{Methodology}
\label{sec:method}

In this section, we present OneRank, a Transformer-native multi-task ranking framework that internalizes multi-task reasoning within the Transformer architecture itself, eliminating the conventional encoder-predictor split $\mathcal{G}(\mathbf{Z}=\mathcal{F}(\mathbf{X}))$. Our design philosophy builds task-private channels alongside task-shared pathways in a bottom-up manner, enabling task specialization while maintaining beneficial knowledge sharing across multiple architectural levels.

We organize our methodology as follows. We first describe how we structure heterogeneous inputs into a unified token sequence representation (\textbf{\S\ref{sec:tokenization}}). To enable early task specialization and mitigate gradient conflicts, we introduce task-specific token injection with mutual invisibility (\textbf{\S\ref{sec:task_encoding}}) that allocates dedicated parameters for each task at the input level. We then design candidate-aware contextualization (\textbf{\S\ref{sec:global_modeling}}) that aggregates cross-candidate signals via situational descriptors, bridging the training-serving gap. To enable controlled cross-task knowledge transfer, we propose flexible cross-task relational attention (\textbf{\S\ref{sec:cross_task_attention}}) with strategic gradient detachment and configurable masking strategies. Finally, we present our joint optimization objectives (\textbf{\S\ref{sec:objectives}}). We elaborate on each component in the following subsections.

\subsection{Structured Tokenization}
\label{sec:tokenization}

Following established paradigms~\cite{dai2025onepiece}, we adopt a structured tokenization strategy to organize heterogeneous inputs. Our approach transforms diverse input modalities into a unified token sequence representation, enabling effective joint modeling of sequential patterns and feature interactions.

\textbf{Interaction History (IH).} 
We organize user behavioral sequences in temporal order as $\mathcal{H} = \{h_1, h_2, \ldots, h_T\}$, where $h_t$ represents the user's interaction at timestamp $t$. To capture temporal dynamics and evolving preferences, we augment each interaction with learnable positional encodings $\mathbf{p}_t \in \mathbb{R}^d$:
$\mathbf{e}_t^{\text{IH}} = \text{Embed}(h_t) + \mathbf{p}_t$,
where $\text{Embed}(\cdot)$ denotes the embedding function and $d$ is the embedding dimension. The complete interaction history sequence is denoted as $\mathcal{S}_{\text{IH}} = [\mathbf{e}_1^{\text{IH}}, \ldots, \mathbf{e}_T^{\text{IH}}]$.

\textbf{Preference Anchoring (PA).}
Inspired by retrieval-augmented generation (RAG) in large language models~\cite{zhao2026retrieval,gao2023retrieval,gupta2024comprehensive}, we enhance user interaction history with external knowledge. Specifically, we introduce Preference Anchors comprising multiple retrieved sequences $\mathcal{A} = \{\mathcal{A}_1, \mathcal{A}_2, \ldots, \mathcal{A}_M\}$ dynamically selected based on domain knowledge. For personalized search, we retrieve top-clicked and top-purchased item sequences related to the current query; for recommendation, we select historically high-engagement sequences as complementary signals. Each sequence $\mathcal{A}_i$ is encapsulated using learnable boundary tokens:
\begin{equation}
\mathcal{S}_{\text{PA}} = \bigoplus_{i=1}^{M} \left(\langle\text{BOS}\rangle \oplus \mathcal{A}_i \oplus \langle\text{EOS}\rangle\right),
\end{equation}
where $\oplus$ denotes concatenation and $M$ is the number of retrieved sequences. $\langle\text{BOS}\rangle$ and $\langle\text{EOS}\rangle$ are learnable tokens representing the beginning and end of each sequence, respectively.

\textbf{Candidate-Task Token Groups.}
For each candidate item $c_i$ in the candidate set $\mathcal{C} = \{c_1, c_2, \ldots, c_N\}$, we construct a token group that includes the candidate embedding $\mathbf{e}_i^{\text{C}} \in \mathbb{R}^d$ and $K$ task-specific tokens. 
The task tokens $\{\mathbf{t}_k\}_{k=1}^K$ are learnable parameters shared across all candidates, where each $\mathbf{t}_k \in \mathbb{R}^d$ serves as a task-specific query template for task $k$. 
For each candidate $c_i$, we instantiate this shared set of task tokens, forming a candidate-task group:
\begin{equation}
\mathcal{G}_i = [\mathbf{e}_i^{\text{C}}, \mathbf{t}_1, \mathbf{t}_2, \ldots, \mathbf{t}_K] \in \mathbb{R}^{(1+K) \times d}.
\end{equation}
Note that while the task token parameters $\{\mathbf{t}_k\}_{k=1}^K$ are shared across all candidate groups, each group operates independently during encoding through structured attention masking, allowing task tokens to extract candidate-specific task representations through attention to different candidate embeddings $\mathbf{e}_i^{\text{C}}$ and the shared user context.

\textbf{Unified Token Sequence.}
The final input concatenates the shared user context followed by all candidate-task groups:
\begin{equation}
\mathcal{X}_0 = [\mathcal{S}_{\text{IH}}, \mathcal{S}_{\text{PA}}, \mathcal{G}_1, \mathcal{G}_2, \ldots, \mathcal{G}_N] \in \mathbb{R}^{S \times d},
\end{equation}
where $S = T + \sum_{i=1}^{M}(|\mathcal{A}_i|+2) + N \cdot (1+K)$ represents the total length. This organization naturally separates task-shared pathways (user context) from task-private channels (task-specific tokens), facilitating efficient attention masking and parallel computation.

\begin{figure*}
    \centering
    \includegraphics[width=0.95\textwidth]{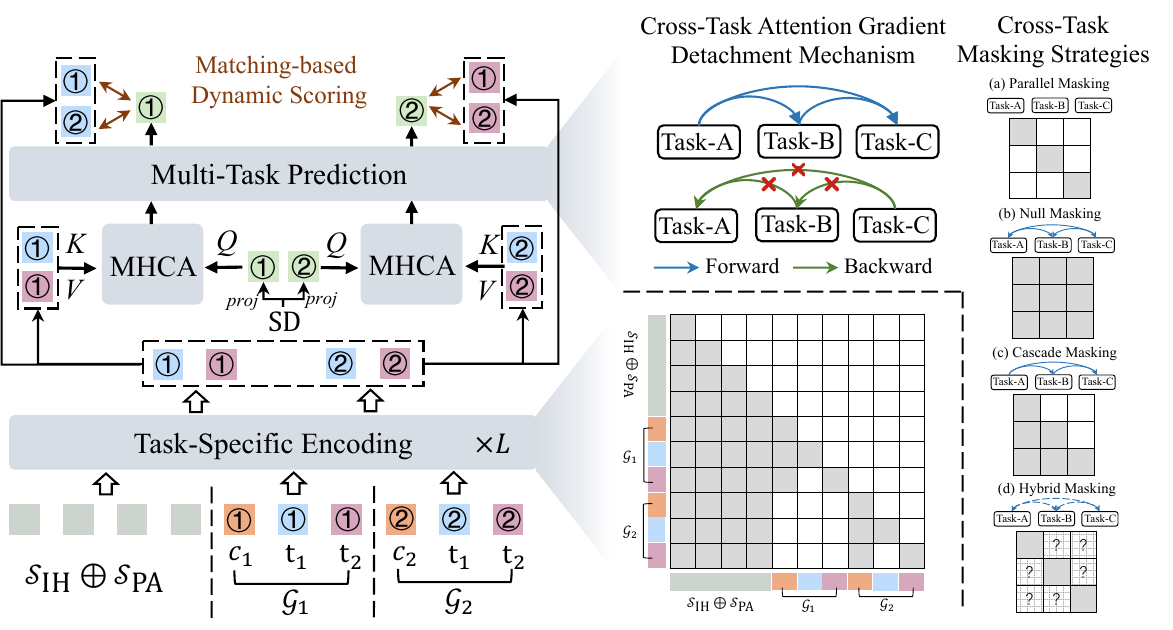}
    \vspace{-15pt}
    \caption{Overall architecture of OneRank. The input is organized into a unified token sequence with structured tokenization, including interaction history (IH), preference anchoring (PA), and candidate-task token groups. Task-specific token injection with mutual invisibility enables early task specialization. Candidate-aware contextualization aggregates cross-candidate signals via situational descriptors, while flexible cross-task relational attention facilitates controlled knowledge transfer across tasks. OneRank adopts a matching-based scoring formulation for dynamic context-aware personalized ranking.}
    \vspace{-10pt}
\end{figure*}

\subsection{Task-Specific Encoding}
\label{sec:task_encoding}

To enable early task specialization and mitigate the seesaw phenomenon in shared-bottom architectures, we introduce task-specific token injection with mutual invisibility at the input level. Unlike conventional approaches that rely on shared representations $\mathbf{Z} = \mathcal{F}(\mathbf{X})$ for all tasks, our design injects shared task-specific token templates into each candidate group, enabling independent task-specific feature extraction through structured attention mechanisms.

\textbf{Structured Attention Mask.}
To ensure task-specific representation learning, we construct a structured attention mask $\mathbf{M} \in \{0,1\}^{S \times S}$ that enforces mutual invisibility among task tokens while maintaining shared user context visibility:

\begin{itemize}[leftmargin=*, topsep=2pt]
    \item \textit{Causal User Context}: Tokens in the user context ($\mathcal{S}_{\text{IH}}$ and $\mathcal{S}_{\text{PA}}$) follow causal attention for temporal modeling, where each position can only attend to itself and preceding positions.
    
    \item \textit{Candidate Group Isolation}: Each group $\mathcal{G}_i$ is isolated from other groups $\mathcal{G}_j$ where $j \neq i$, enabling efficient single-user multiple-candidate parallelization. Tokens within $\mathcal{G}_i$ can attend to:
    \begin{itemize}
        \item All tokens in the user context with causal masking
        \item The candidate embedding $\mathbf{e}_i^{\text{C}}$ within the same group
        \item Themselves (self-attention)
    \end{itemize}
    
    \item \textit{Task Token Mutual Invisibility}: Task tokens from different tasks are mutually invisible even within the same candidate group. Specifically, the $k$-th task token in group $\mathcal{G}_i$ can only attend to the user context (with causal masking), the candidate embedding $\mathbf{e}_i^{\text{C}}$, and itself, but cannot see other task tokens in the same group.
\end{itemize}

Formally, let $\text{pos}(p)$ denote the sequential position of token $p$ in the user context, and let $\mathbf{t}_k^{(i)}$ denote the $k$-th task token in candidate group $\mathcal{G}_i$ (instantiated from the shared parameter $\mathbf{t}_k$). For token positions $p$ and $q$, the mask is defined as:
\begin{equation}
\mathbf{M}_{pq} = 
\begin{cases}
1, & \text{if } p, q \in \{\mathcal{S}_{\text{IH}}, \mathcal{S}_{\text{PA}}\} \text{ and } \text{pos}(q) \leq \text{pos}(p) \\
1, & \text{if } p \in \mathcal{G}_i \text{ and } q \in \{\mathcal{S}_{\text{IH}}, \mathcal{S}_{\text{PA}}\} \\
1, & \text{if } p \in \mathcal{G}_i \text{ and } q = \mathbf{e}_i^{\text{C}} \\
1, & \text{if } p = \mathbf{t}_k^{(i)} \text{ and } q = \mathbf{t}_k^{(i)} \\
0, & \text{otherwise}
\end{cases}
\end{equation}

\textbf{Transformer Encoding.}
We apply $L$ layers of masked \textit{Multi-Head Self-Attention (MHSA)} with residual connections and layer normalization for better training stability:
\begin{equation}
\begin{aligned}
\mathbf{H}^{(\ell)} &= \text{LN}\left(\text{MHSA}^{(\ell)}(\mathcal{X}^{(\ell-1)}, \mathbf{M})\right) + \mathcal{X}^{(\ell-1)}, \\
\mathcal{X}^{(\ell)} &= \text{LN}\left(\text{FFN}^{(\ell)}(\mathbf{H}^{(\ell)})\right) + \mathbf{H}^{(\ell)},
\end{aligned}
\end{equation}
where $\ell \in \{1, \ldots, L\}$ indexes the layer. After encoding, we extract task-specific representations by selecting the output of the corresponding task token from each candidate group:
\begin{equation}
\mathbf{r}_k^i = \text{Extract}(\mathcal{X}^{(L)}, \mathbf{t}_k^{(i)}) \in \mathbb{R}^d,
\end{equation}
where $\mathbf{r}_k^i$ encodes task-relevant features for candidate $i$ in task $k$. Although all candidate groups share the same task token parameters $\{\mathbf{t}_k\}_{k=1}^K$, each task token produces different representations $\mathbf{r}_k^i$ by attending to different candidate embeddings $\mathbf{e}_i^{\text{C}}$ and integrating candidate-specific signals from the shared user context, achieving early task specialization.

\subsection{Candidate-Aware Contextualization}
\label{sec:global_modeling}

Traditional point-wise scoring suffers from a training-serving gap: models trained on isolated samples fail to capture cross-candidate dependencies present during serving. We address this through candidate-aware contextualization that aggregates cross-candidate signals via situational descriptors.

\textbf{Situational Descriptors (SD).}
We define a Situational Descriptor $\mathbf{s} \in \mathbb{R}^{d}$ that encapsulates contextual signals including user demographics, query information, and session metadata (\textit{e.g.}, time, location). This serves as a contextual anchor for aggregation.

\textbf{Task-Specific Cross-Candidate Aggregation.}
For each task $k$, we employ task-specific parameters to transform the SD and aggregate candidate information. Specifically, we use a task-specific projection function $f_k(\cdot)$ to transform the situational descriptor:
\begin{equation}
\mathbf{q}_k = \text{LN}(f_k(\mathbf{s})) \in \mathbb{R}^d,
\end{equation}
where $f_k(\cdot)$ is a learnable projection with independent parameters for each task. We then employ task-specific \textit{Multi-Head Cross-Attention (MHCA)}~\cite{vaswani2017attention} to aggregate candidate-aware global information:
\begin{equation}
\mathbf{h}_k = \text{MHCA}_k\left(\mathbf{q}_k, \{\mathbf{r}_k^i\}_{i=1}^N, \{\mathbf{r}_k^i\}_{i=1}^N\right) \in \mathbb{R}^d,
\end{equation}
where $\text{MHCA}_k(\cdot)$ denotes task-specific multi-head cross-attention with dedicated parameters for task $k$, and $\mathbf{h}_k$ represents the task-wise global representation for task $k$, aggregated over the entire candidate set. This design explicitly decouples task-specific information flows through independent parameter sets, ensuring that each task maintains its own aggregation pathway while capturing cross-candidate competitive dynamics.

\subsection{Multi-Task Prediction}
\label{sec:cross_task_attention}

To enable controlled cross-task knowledge transfer while respecting domain-specific dependencies, we design flexible cross-task relational attention with strategic gradient detachment. Unlike conventional approaches that employ fixed task tower structures, our framework allows configurable information flow patterns.

\textbf{Cross-Task Attention with Strategic Gradient Detachment.}
We organize task representations $\{\mathbf{h}_k\}_{k=1}^K$ obtained from candidate-aware contextualization and apply multi-head self-attention with a configurable cross-task attention mask $\mathbf{A} \in \{0,1\}^{K \times K}$:
\begin{equation}
\tilde{\mathbf{h}}_k = \text{MHSA}\left(\mathbf{h}_k, \{\mathbf{h}_j\}_{j: \mathbf{A}_{kj}=1}\right),
\end{equation}
where task $k$ attends only to tasks $j$ where $\mathbf{A}_{kj}=1$. 

To prevent backward gradient interference while allowing forward knowledge transfer, we employ strategic gradient detachment. Specifically, we customize the backward operator of the cross-task attention to only allow diagonal gradient flow while blocking off-diagonal gradients. During backpropagation, when computing gradients for task $k$, we detach gradients from attended tasks $j \neq k$ by setting $\frac{\partial \mathcal{L}}{\partial \mathbf{h}_j} = 0$ for $j \neq k$ in the attention computation. This ensures that optimizing task $k$ does not adversely affect the learning of other tasks, effectively mitigating inter-task gradient conflicts while preserving beneficial forward information transfer, turning cross-task attention into a read-only memory for knowledge transfer. We then apply residual connection and layer normalization:
\begin{equation}
\hat{\mathbf{h}}_k = \text{LN}(\tilde{\mathbf{h}}_k) + \mathbf{h}_k.
\end{equation}

\textbf{Dynamic Matching-Based Scoring.}
We refine representations through a feed-forward network with residual connection:
\begin{equation}
\mathbf{z}_k = \text{LN}(\text{FFN}(\hat{\mathbf{h}}_k)) + \hat{\mathbf{h}}_k \in \mathbb{R}^d.
\end{equation}
Unlike static MLP-based scoring that applies fixed transformations regardless of context, we compute task-candidate relevance through inner product similarity:
\begin{equation}
s_k^i = \mathbf{z}_k^\top \mathbf{r}_k^i,
\end{equation}
where $\mathbf{z}_k$ (enriched by controlled cross-task interactions) captures task-aware global context, and $\mathbf{r}_k^i$ (from task-specific encoding) captures context-conditioned candidate embeddings. This Transformer-native matching formulation adapts dynamically to session context, enabling context-aware and task-adaptive ranking.

\textbf{Flexible Cross-Task Masking Strategies.}
The cross-task attention mask $\mathbf{A}$ can be flexibly configured based on domain-specific task relationships and expert knowledge. We discuss several representative strategies below:
\begin{itemize}[leftmargin=*]
    \item \textit{\textbf{Parallel Masking}}: When independent predictions among different recommendation tasks are desired, we enforce mutual invisibility ($\mathbf{A}_{kj} = \mathbb{I}[k=j]$). Each task relies solely on its own global representation without cross-task information flow. This strategy is suitable for exploratory scenarios or when data abundance allows independent modeling.
    
    \item \textit{\textbf{Null Masking}}: For scenarios with abundant data where task relationships are complex and ambiguous, we allow all tasks to attend to each other ($\mathbf{A}_{kj} = 1, \forall k,j$). The model autonomously learns task correlations through the bidirectional attention, suitable when domain knowledge about task dependencies is limited.
    
    \item \textit{\textbf{Cascade Masking}}: When behavioral dependencies follow a clear funnel structure, we impose unidirectional information flow following a data-rich-to-sparse cascade ($\mathbf{A}_{kj} = \mathbb{I}[j \leq k]$). This enables sparse downstream tasks to leverage signals from upstream tasks. For instance, in e-commerce, the natural progression \textit{click} $\rightarrow$ \textit{cart} $\rightarrow$ \textit{purchase} exhibits clear causal dependencies, where purchase prediction benefits from click and cart signals.
    
    \item \textit{\textbf{Hybrid Masking}}: For complex real-world scenarios, practitioners can design custom masks encoding partial visibility or mixed patterns based on domain expertise. For example, in short-video platforms where behavioral relationships among \textit{like}, \textit{follow}, \textit{comment}, and \textit{forward} lack clear causal structure, one might allow bidirectional attention between engagement-related tasks (\textit{like}, \textit{comment}) while maintaining unidirectional flow from abundant \textit{click} signals to sparse \textit{follow} actions.
\end{itemize}
In summary, our flexible masking mechanism enables controlled cross-task interaction based on domain-specific task dependencies, accommodating diverse relationship patterns from strict cascades to fully autonomous learning.

\subsection{Joint Learning Objectives}
\label{sec:objectives}

With relevance scores $s_k^i = \mathbf{z}_k^\top \mathbf{r}_k^i$ computed via the dynamic matching mechanism (\textbf{\S\ref{sec:cross_task_attention}}), we employ a hybrid learning strategy combining list-wise and point-wise objectives.
For discriminative ranking, we adopt InfoNCE-based contrastive loss~\cite{rusak2024infonce,yi2025recgpt,tang2024towards}:
\begin{equation}
\label{eq:loss_infonce}
\mathcal{L}_k^{\text{list}} = -\sum_{i \in \mathcal{I}_k^+} \log \frac{\exp(s_k^i/\tau)}{\sum_{j=1}^N \exp(s_k^j/\tau)},
\end{equation}
where $\mathcal{I}_k^+$ denotes positive samples, $\tau$ is temperature, and $N$ is candidate set size. For calibrated probability estimation required in industrial systems, we employ binary cross-entropy (BCE) loss:
\begin{equation}
\mathcal{L}_k^{\text{point}} = -\sum_{i=1}^N \left[y_k^i \log \sigma(s_k^i) + (1-y_k^i) \log(1-\sigma(s_k^i))\right],
\end{equation}
where $y_k^i \in \{0,1\}$ is the ground-truth label and $\sigma(\cdot)$ is the sigmoid function. The joint training objective combines both losses across all tasks, formulated as:
\begin{equation}
\label{eq:final_loss}
\mathcal{L} = \sum_{k=1}^K \left(\alpha \mathcal{L}_k^{\text{list}} + \beta \mathcal{L}_k^{\text{point}}\right),
\end{equation}
where $\alpha$ and $\beta$ balance list-wise and point-wise optimization.

\section{Discussion}
\label{sec:discussion}

Our unified framework offers several fundamental advantages over conventional $\mathcal{F}$-$\mathcal{G}$ decoupled approaches. We organize the discussion around three core design principles.

\subsection{\textbf{Bridging the Training-Serving Gap via Context-Aware Dynamic Ranking}}

Traditional point-wise learning paradigms~\cite{xin2022prototype,yang2022cross,lin2022personalized} optimize individual user-item pairs in isolation, creating a fundamental mismatch with the serving environment where models must rank entire candidate sets. OneRank addresses this gap through integrated context-aware modeling and dynamic scoring.

\textbf{Explicit Cross-Candidate Dependency Modeling.} Our situational descriptor-based global information modeling (\textbf{\S\ref{sec:global_modeling}}) aggregates signals across the entire candidate set via cross-attention, enabling the model to capture competitive dynamics and relative preferences rather than absolute scores. The global representation $\mathbf{h}_k$ encodes not only task-specific characteristics but also the distributional properties of the candidate pool, allowing the model to adaptively adjust rankings based on the competing items.

\textbf{Dynamic Ranking Scoring.} Unlike static MLP predictors that apply fixed transformations regardless of context, our matching-based formulation $s_k^i = \mathbf{z}_k^\top \mathbf{r}_k^i$ enables dynamic adaptation. The global representation $\mathbf{z}_k$, refined through cross-task attention and informed by situational descriptors, captures session-specific user intent, query semantics, and temporal context. Consequently, the same user-item pair can receive different scores across sessions based on contextual variations (\textit{e.g.}, morning \textit{vs.} evening browsing, search \textit{vs.} browse mode), achieving true personalized ranking. Furthermore, the inner product formulation induces a shared geometric space where task representations $\{\mathbf{z}_k\}$ and candidate representations $\{\mathbf{r}_k^i\}$ are jointly optimized for semantic alignment, facilitating better gradient flow and more effective multi-task learning compared to architecturally separated MLP towers.

\begin{table*}[t]
\centering
\caption{Offline performance comparison under different encoder architectures and multi-task learning strategies, together with model size (Params) and computational cost (FLOPs). Best results are highlighted in bold.}
\vspace{-10pt}
\label{tab:offline_main}
\resizebox{0.9 \linewidth}{!}{
\large
\begin{tabular}{llcccccccc}
\toprule
\textbf{Encoder} & \textbf{Predictor} 
& \textbf{Params} & \textbf{FLOPs}
& \textbf{C-AUC} & \textbf{C-GAUC} & \textbf{A-AUC} & \textbf{A-GAUC} & \textbf{O-AUC} & \textbf{O-GAUC} \\
\midrule
\multirow{6}{*}{DNN (RecSys'16)}
& noMTL 
& 9.5M & 226.7M
& 0.7638 & 0.7664 & 0.8193 & 0.7813 & 0.8824 & 0.8128 \\
& NSE 
& 6.7M & 161.5M
& 0.7667 & 0.7693 & 0.8242 & 0.7863 & 0.8859 & 0.8181 \\
& MMoE (KDD'18) 
& 8.5M & 235.1M
& 0.7682 & 0.7723 & 0.8246 & 0.7893 & 0.8881 & 0.8223 \\
& PLE (RecSys'20) 
& 9.2M & 220.6M
& 0.7688 & 0.7721 & 0.8277 & 0.7906 & 0.8872 & 0.8198 \\
& DCMT (ICDE'23) 
& 9.6M & 302.3M
& 0.7642 & 0.7669 & 0.8203 & 0.7817 & 0.8828 & 0.8125 \\
& ResFlow (KDD'24) 
& 9.5M & 226.7M
& 0.7649 & 0.7688 & 0.8252 & 0.7886 & 0.8886 & 0.8204 \\
\midrule
\multirow{5}{*}{MTGR (CIKM'25)}
& NSE 
& 2.3M & 487.0M
& 0.7720 & 0.7732 & 0.8304 & 0.7910 & 0.8924 & 0.8240 \\
& MMoE (KDD'18) 
& 2.3M & 487.0M
& 0.7718 & 0.7737 & 0.8304 & 0.7915 & 0.8914 & 0.8237 \\
& PLE (RecSys'20) 
& 2.7M & 496.0M
& 0.7723 & 0.7737 & 0.8318 & 0.7930 & 0.8933 & 0.8260 \\
& DCMT (ICDE'23) 
& 2.8M & 497.7M
& 0.7208 & 0.7282 & 0.7888 & 0.7583 & 0.8629 & 0.7986 \\
& ResFlow (KDD'24) 
& 2.7M & 500.1M
& 0.7703 & 0.7717 & 0.8294 & 0.7901 & 0.8919 & 0.8253 \\
\midrule
\multirow{5}{*}{OneTrans (WWW'26)}
& NSE 
& 6.3M & 823.2M
& 0.7773 & 0.7772 & 0.8359 & 0.7961 & 0.8991 & 0.8280 \\
& MMoE (KDD'18) 
& 6.1M & 822.7M
& 0.7768 & 0.7765 & 0.8364 & 0.7934 & 0.8973 & 0.8241 \\
& PLE (RecSys'20) 
& 6.4M & 823.5M
& 0.7770 & 0.7775 & 0.8371 & 0.7982 & 0.8996 & 0.8336 \\
& DCMT (ICDE'23) 
& 6.4M & 823.4M
& 0.7574 & 0.7596 & 0.8141 & 0.7746 & 0.8869 & 0.8149 \\
& ResFlow (KDD'24) 
& 6.3M & 823.2M
& 0.7764 & 0.7766 & 0.8372 & 0.7975 & 0.8987 & 0.8287 \\
\midrule
\multicolumn{2}{c}{\textbf{OneRank (Ours)}}
& 4.9M & 1.0G
& \textbf{0.7910} & \textbf{0.7843} 
& \textbf{0.8463} & \textbf{0.8036} 
& \textbf{0.9024} & \textbf{0.8350} \\
\bottomrule
\end{tabular}
}
\end{table*}

\subsection{\textbf{Mitigating the Seesaw Phenomenon via Decoupled Optimization}}

The seesaw phenomenon, where optimizing one task degrades others, arises from gradient conflicts on shared parameters in multi-task learning. OneRank mitigates this through a three-level decoupling strategy:

\textbf{Input-Level Task-Specific Parameters.} By injecting learnable task tokens $\{\mathbf{t}_k^i\}$ with task-isolated attention masks (\textbf{\S\ref{sec:task_encoding}}), we allocate dedicated parameters for each task at the earliest stage of feature extraction. This ensures that tasks extract specialized representations from the shared context (interaction history, preference anchors) without interfering with each other's gradient flows during backpropagation through the encoding layers.

\textbf{Intermediate-Level Information Flow Decoupling.} At the global modeling stage (\textbf{\S\ref{sec:global_modeling}}), we employ task-specific parameters for both situational descriptor projection ($f_k(\mathbf{s})$) and cross-candidate aggregation ($\text{MHCA}_k$). Even though tasks share the same situational descriptor input $\mathbf{s}$, each task maintains independent transformation and aggregation pathways with dedicated parameter sets. This architectural design prevents gradient conflicts at the aggregation stage: optimizing task $k$'s projection and attention parameters does not directly interfere with other tasks' learning, as each task operates through its own parameter space.

\textbf{Prediction-Level Gradient Detachment.} At the final decoding stage (\textbf{\S\ref{sec:cross_task_attention}}), our gradient detachment mechanism allows task $k$ to benefit from other tasks' representations in the forward pass (knowledge transfer) while preventing its gradients from flowing back to other tasks (optimization isolation). By customizing the backward operator to only allow diagonal gradient flow, we achieve asymmetric information transfer: forward sharing enables knowledge transfer, while backward isolation eliminates negative interference.

\subsection{\textbf{Enhanced Flexibility and Efficiency via Unified Architecture}}

Beyond addressing the seesaw phenomenon and training-serving gap, OneRank's unified design also offers significant advantages in modeling flexibility and computational efficiency.

\textbf{Flexible Task Dependency Modeling.} Our configurable cross-task attention masks (\textbf{\S\ref{sec:cross_task_attention}}) provide unprecedented flexibility compared to rigid architectures like ESMM's fixed cascades~\cite{ma2018entire} or MMoE's independent towers~\cite{ma2018modeling}. Practitioners can encode domain-specific knowledge through simple mask design: strict cascades for e-commerce funnels, bidirectional attention for ambiguous engagement patterns, or hybrid configurations for complex user journeys. This flexibility eliminates the need for architecture search or task-specific model variants, enabling rapid adaptation to diverse scenarios within a single unified framework.

\textbf{Computational Efficiency and Scalability.} By eliminating the $\mathcal{F}$-$\mathcal{G}$ transition, OneRank achieves end-to-end optimization within a single Transformer-native architecture, avoiding the computational overhead of heterogeneous module transitions present in hybrid approaches. Our single-user multiple-candidate paradigm substantially reduces redundant context encoding during training, while KV-caching of user-specific components (interaction history and preference anchors) enables efficient serving: only candidate and task tokens require online computation, achieving low latency and high GPU utilization. This unified design unlocks the full scaling potential of Transformers, supporting deeper models and larger candidate sets without architectural bottlenecks.

In summary, OneRank's \textbf{co-design} of feature extraction, task-specific representation learning, and multi-task decoding within a unified Transformer architecture addresses the fundamental limitations of conventional decoupled approaches, offering superior modeling capacity, optimization stability, computational efficiency, and adaptability for industrial multi-task ranking systems.

\section{Offline Evaluation}
\label{sec:experiment}

\begin{table}[t]
\centering
\caption{Statistics of the Shopee dataset. Abbreviations: M = Million (10\textsuperscript{6}), B = Billion (10\textsuperscript{9}).}
\vspace{-10pt}
\label{tab:shopee_dataset_stats}
\resizebox{1\linewidth}{!}{
\begin{tabular}{@{}cccccccc@{}}
\toprule
\textbf{\#User} & \textbf{\#Item} & \textbf{\#Query} & \textbf{\#Impression} & \textbf{\#Click} & \textbf{\#Add-to-Cart} & \textbf{\#Order} \\
\midrule
33M & 118M & 105M & 26.6B & 1.05B & 251M & 40M \\
\bottomrule
\end{tabular}
}
\vspace{-15pt}
\end{table}

\subsection{Experimental Setup}

We conduct offline experiments on a large-scale proprietary dataset collected from Shopee, a leading e-commerce platform. The dataset spans 30 consecutive days of user interaction logs in December 2025, covering click, add-to-cart, and order feedback signals. Table~\ref{tab:shopee_dataset_stats} summarizes the dataset statistics. We report AUC and GAUC for click (C), add-to-cart (A), and order (O) prediction tasks. Complete dataset details are provided in Appendix~\ref{appendix:implementation}.

\paragraph{Baseline Methods.}
To provide a comprehensive evaluation, we compare OneRank against combinations of different encoder architectures and multi-task learning strategies. This experimental design allows us to disentangle the impact of encoder capacity from that of multi-task optimization strategies, and to fairly assess the benefits of OneRank's unified Transformer-native multi-task ranking architecture.

We consider the following representative encoder architectures:
\begin{itemize}[leftmargin=0.5cm]
    \item \textbf{DNN}: A well-optimized production baseline in Shopee based on deep neural networks, serving as the foundation for traditional DLRM-style architectures.
    \item \textbf{MTGR}~\cite{han2025mtgr}: An industrial-scale generative recommendation framework developed at Meituan that employs Transformer-based sequence modeling for user behavior understanding.
    \item \textbf{OneTrans}~\cite{zhang2025onetrans}: A unified Transformer architecture for feature interaction and sequential modeling proposed by ByteDance, representing the state-of-the-art in Transformer-based ranking models.
\end{itemize}

On top of each encoder, we evaluate multiple multi-task learning strategies:
\begin{itemize}[leftmargin=0.5cm]
    \item \textbf{noMTL}: Independent single-task training without multi-task learning, serving as a baseline to quantify the benefits of multi-task optimization.
    \item \textbf{NSE (Naive Shared Embedding)}: Separate task-specific networks sharing a common embedding table across tasks, representing the simplest form of parameter sharing.
    \item \textbf{MMoE}~\cite{ma2018modeling}: Multi-gate Mixture-of-Experts that employs task-specific gating networks over shared expert networks, enabling flexible task-specific feature extraction.
    \item \textbf{PLE}~\cite{tang2020progressive}: Progressive Layered Extraction with both shared and task-specific experts organized in a progressive manner to balance knowledge sharing and task specialization.
    \item \textbf{DCMT}~\cite{zhu2023dcmt}: A debiasing-oriented multi-task framework that addresses sample selection bias through causal-based counterfactual learning techniques.
    \item \textbf{ResFlow}~\cite{fu2024residual}: A residual-based multi-task learning approach that enables flexible information flow across tasks through simple residual connections between multi-task towers.
\end{itemize}

\paragraph{Implementation Details.}
Unless otherwise specified, all models are implemented under the same experimental settings to ensure fair comparison.
OneRank employs a 2-layer Transformer encoder with pre-norm architecture~\cite{yang2025qwen3,team2024qwen2,liu2024deepseek} and 4 attention heads per layer. The model uses a hidden dimension of 256 throughout, with feed-forward networks set to twice the hidden size, following standard Transformer configurations. The maximum sequence length is set to 256 to balance computational efficiency and modeling capacity. Learnable positional encodings~\cite{tang2025think,tang2026parallel} are applied to capture temporal dependencies in user interaction sequences.

Sequence-side features (including item attributes, category information, and behavioral metadata) and situational descriptors (user demographics, query information, session metadata) are projected to the model dimension (256) via linear layers. OneRank instantiates three task-specific tokens (one for each task: click, add-to-cart, order) and three ranking head tokens. The multi-task prediction module applies one layer of task-specific self-attention followed by one layer of cross-task attention with configurable masking, both equipped with FFN modules for non-linear transformation.

For the InfoNCE loss (Eq.~\eqref{eq:loss_infonce}), the initial temperature $\tau$ is set to 0.2. The list-wise and point-wise losses are weighted equally ($\alpha = \beta = 1$ in Eq.~\eqref{eq:final_loss}), and all tasks are assigned uniform loss weights.

\begin{figure}[t]
    \centering
    \includegraphics[width=\linewidth]{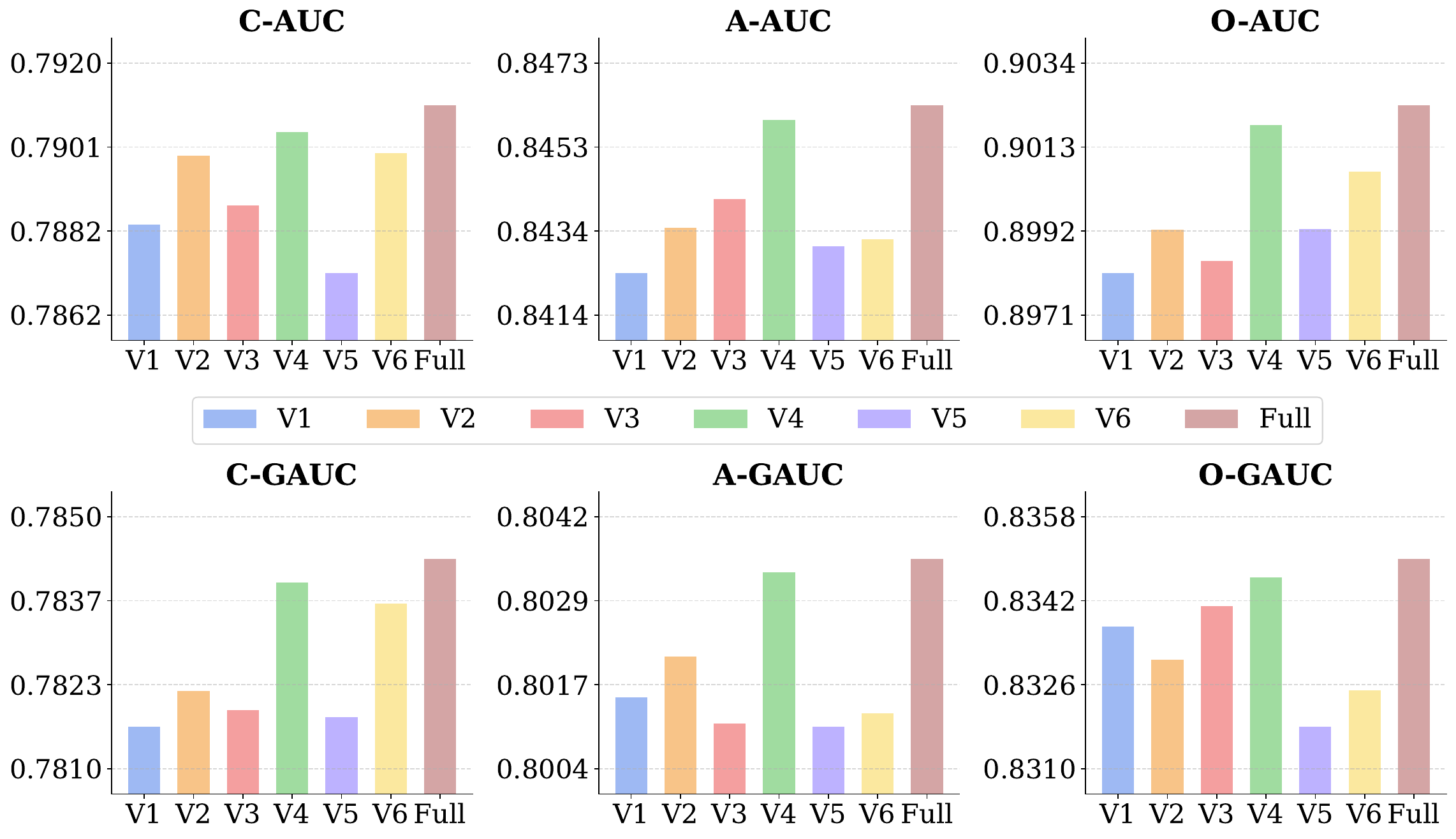}
    \vspace{-20pt}
    \caption{Ablation studies.}
    \label{fig:ablation}
    \vspace{-15pt}
\end{figure}

\subsection{Overall Performance}
Table~\ref{tab:offline_main} summarizes the overall offline performance, from which three key observations can be drawn.

\textbf{Multi-task learning is consistently beneficial under conventional DNN encoders.}
Compared with the noMTL baseline, all multi-task learning methods yield clear improvements across click, add-to-cart, and order prediction tasks when built on top of a DNN encoder. This result confirms that jointly modeling dense and sparse user feedback provides complementary supervision signals, and establishes multi-task learning as a necessary component for industrial recommender systems.

\textbf{Stronger encoder architectures further amplify the benefits of multi-task modeling.}
Replacing DNN-based encoders with Transformer-based architectures such as MTGR and OneTrans leads to additional performance gains across most multi-task strategies, highlighting the importance of expressive sequence and context modeling in multi-task recommendation. However, we observe that DCMT performs poorly under Transformer encoders, likely because its debiasing-oriented design may over-correct sparse tasks and exacerbate task imbalance, resulting in unstable optimization when combined with high-capacity models.

\textbf{OneRank achieves the best performance by unifying representation learning and multi-task ranking within a Transformer-native framework.}
Across all metrics and tasks, OneRank consistently outperforms all baseline combinations, demonstrating the effectiveness of eliminating the encoder--predictor decoupling and internalizing multi-task reasoning directly within the Transformer architecture. Notably, these gains are achieved with a compact parameterization and a moderate increase in computation, highlighting a favorable performance–efficiency trade-off. These results validate our central claim that a unified, Transformer-native ranking paradigm is more suitable for large-scale multi-task recommendation than externally encoder-predictor separation architectures.

\subsection{Ablation Studies}

To validate the effectiveness of each design component in OneRank, we conduct comprehensive ablation studies by removing or replacing key architectural modules. Figure~\ref{fig:ablation} presents the results of the following variants: \textbf{V1} removes task-specific tokens and directly applies linear projections on candidate encodings; \textbf{V2} replaces $K$ task-specific tokens with a single shared token; \textbf{V3} further removes cross-task relational attention on top of V2; \textbf{V4} removes strategic gradient detachment in cross-task attention; \textbf{V5} replaces situational descriptors with randomly initialized parameters; \textbf{V6} applies full bidirectional attention without selective masking.

From the results, we observe that: \textbf{(1)} Removing task-specific tokens (V1) leads to notable degradation, with A-AUC dropping from 0.8463 to 0.8424 and O-GAUC declining from 0.8350 to 0.8337, validating the necessity of early task specialization; \textbf{(2)} Using a single shared token (V2) underperforms the full model, confirming that independent task tokens are crucial for mitigating gradient conflicts; \textbf{(3)} Removing cross-task attention (V3) shows mixed results compared to V2, suggesting that cross-task knowledge transfer benefits when combined with proper task isolation; \textbf{(4)} Removing gradient detachment (V4) achieves strong performance but causes instability on add-to-cart (A-AUC: 0.8460 vs. 0.8463), demonstrating the necessity of strategic gradient control; \textbf{(5)} Replacing situational descriptors with random parameters (V5) causes severe degradation with C-AUC dropping to 0.7872 and O-GAUC to 0.8318, confirming that candidate-aware contextualization is essential for bridging the training-serving gap; \textbf{(6)} Full bidirectional masking (V6) consistently underperforms controlled masking strategies across all tasks. Overall, these studies validate that the synergy of all proposed components is essential for superior multi-task ranking performance.

\begin{figure}[t]
    \centering
    \includegraphics[width=\linewidth]{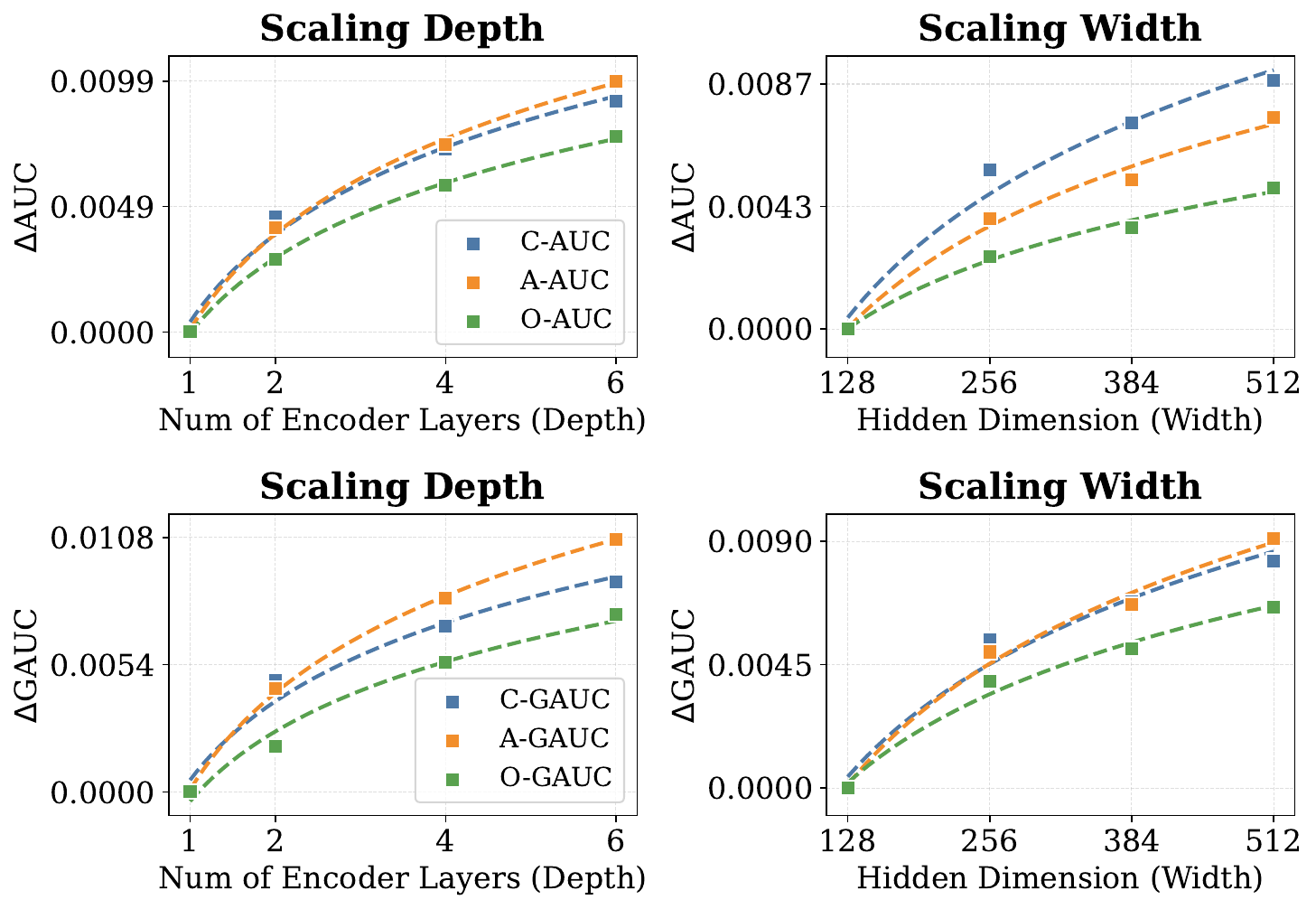}
    \vspace{-15pt}
    \caption{Scaling performance analysis of OneRank w.r.t. encoder depth and hidden size.}
    \label{fig:scale_delta_auc_gauc_2x2}
    \vspace{-20pt}
\end{figure}

\subsection{Scaling Analysis}

To investigate the scaling behavior of OneRank, we study its performance under two fundamental model scaling dimensions: the \textbf{encoder layer (depth)} and the \textbf{hidden dimension (width)}. Specifically, we progressively increase the number of encoder layers and the model dimensionality, while keeping all other components unchanged. We report absolute performance improvements in terms of $\Delta$AUC and $\Delta$GAUC \textit{w.r.t.} the smallest model configuration.

\textbf{Scaling Encoder Depth.}
The left column of Figure~\ref{fig:scale_delta_auc_gauc_2x2} shows the effect of increasing encoder depth. We observe consistent performance improvements across all feedback signals as the number of encoder layers increases, for both AUC and GAUC metrics. The gains are most pronounced when scaling from shallow configurations to moderate depths, and gradually saturate at larger depths, indicating diminishing marginal returns. This trend suggests that deeper Transformer stacks enhance OneRank’s capacity to model complex sequential patterns and task interactions, while the unified architecture maintains stable optimization as depth increases.

\textbf{Scaling Hidden dimension.}
The right column of Figure~\ref{fig:scale_delta_auc_gauc_2x2} illustrates the impact of scaling the hidden dimension. Increasing the model width consistently improves ranking performance across different tasks. Compared with depth scaling, width scaling exhibits smoother and more uniform gains, particularly for click-related metrics, reflecting the benefit of richer representation capacity for fine-grained user–item interaction modeling. Notably, performance improvements remain stable across tasks, indicating that OneRank effectively leverages increased model capacity without introducing severe task imbalance or optimization instability.

Overall, the consistent improvements across multiple tasks and metrics further support the scalability of the proposed unified Transformer-native ranking framework, making it well suited for large-scale industrial deployment where model capacity and performance must be jointly optimized.

\subsection{Online A/B Testing}

To validate the practical effectiveness of OneRank in real-world industrial environments, we conduct large-scale online A/B testing on Shopee's main personalized ranking scenario over a 7-day period, comparing against a previously deployed baseline that combines a carefully optimized Transformer encoder with a multi-task predictor. OneRank is fully deployed within the standard multi-stage ranking pipeline with score fusion to balance user experience and business objectives. Following industrial practices, we allocate 10\% of live traffic to the treatment group using OneRank and 10\% to a baseline control group. We evaluate online performance along two complementary dimensions: \textbf{Platform Benefits} (GMV/UU, Paid GMV/UU, AR/UU) and \textbf{User Experience} (Bad Query Rate). Detailed deployment configurations are provided in Appendix~\ref{appendix:online}.

The online A/B testing results are summarized in Table~\ref{tab:online_ab}. OneRank consistently improves both business-centric and user-centric metrics, achieving a \textbf{+1.01\%} lift in GMV/UU, \textbf{+1.17\%} increase in paid GMV per user, and \textbf{+0.81\%} gain in advertising revenue, while simultaneously reducing Bad Query Rate by \textbf{2.29\%}. These results demonstrate that OneRank not only enhances monetization efficiency but also improves recommendation relevance and user experience in industrial scenarios. Overall, the observed gains validate the practical effectiveness of OneRank, and highlight the strong potential of unified Transformer-native designs for scalable and reliable deployment in large-scale industrial recommender systems.

\begin{table}[t]
\centering
\caption{Online A/B testing results (relative improvement) comparing OneRank with the production baseline.}
\vspace{-5pt}
\label{tab:online_ab}
\resizebox{0.8 \linewidth}{!}{
\large
\begin{tabular}{cccc}
\toprule
GMV/UU$\uparrow$ & Paid GMV/UU$\uparrow$ & AR/UU$\uparrow$ & Bad Query Rate$\downarrow$ \\
\midrule
\textbf{+1.01\%} & \textbf{+1.17\%} & \textbf{+0.81\%} & \textbf{-2.29\%} \\
\bottomrule
\end{tabular}
}
\vspace{-10pt}
\end{table}

\section{Related Work}
\label{sec:related_work}

Our work is closely related to multi-task learning in recommender systems and Transformer-based ranking architectures. We briefly review representative work.

\paragraph{\textbf{Multi-Task Learning for Recommendation.}}
Multi-task learning has become essential in modern recommender systems to jointly model diverse user behaviors~\cite{bai2022contrastive,xu2022mixture,qin2020multitask,li2020multi}. Existing approaches primarily exploit task dependencies through two paradigms. The first focuses on \emph{explicit dependency modeling} through structured knowledge transfer. ESMM~\cite{ma2018entire} exploits the conditional relationship between CTR and CVR to address data sparsity, while ESCM~\cite{wang2022escm2} further refines this with counterfactual reasoning. AITM~\cite{xi2021modeling} and ResFlow~\cite{fu2024residual} extend this paradigm with attention-based adaptive transfer and residual connections, respectively. However, these methods rely on predefined task structures or heuristic transfer rules, limiting adaptability. The second paradigm focuses on \emph{implicit knowledge sharing} through dynamic routing mechanisms. MMoE~\cite{ma2018modeling} introduces mixture-of-experts with task-specific gating, while SNR~\cite{ma2019snr} and PLE~\cite{tang2020progressive} propose progressive layered extraction with shared and task-specific experts. OnePiece~\cite{dai2025onepiece} ranking model further connects multiple tasks through reasoning tokens, but still follows an encoder-predictor separation with MLP-based scoring.
Despite their flexibility, these approaches compress features into task-agnostic shared representations, creating information bottlenecks and gradient conflicts, while lacking inherent scaling mechanisms for large-scale deployments.

\paragraph{\textbf{Transformer-Based Ranking.}}
Inspired by large language models~\cite{naveed2025comprehensive,zhao2023survey,zhang2026instruction}, recent work has explored Transformer for ranking~\cite{gui2023hiformer,guan2025make,huang2026hyformer,yu2025hhft,chen2025homer,shenqiang2026gap,lai2026unleashing,zhang2024wukong,zhai2024actions,zeng2024interformer,tang2026loopctr}. KuaiFormer~\cite{liu2024kuaiformer} focuses on long-sequence modeling, while Climber~\cite{xu2025climber} addresses heterogeneous sequences. Unified architectures including HHFT~\cite{yu2025hhft}, MTGR~\cite{han2025mtgr}, OneTrans~\cite{zhang2025onetrans}, and HyFormer~\cite{huang2026hyformer} jointly model feature interactions and sequential patterns within a single Transformer framework. However, these methods retain the conventional encoder-predictor paradigm $\mathcal{G}(\mathbf{Z}=\mathcal{F}(\mathbf{X}))$, where a Transformer encoder produces task-agnostic representations $\mathbf{Z}$ fed into MLP-based task towers. This design introduces three limitations: (1) the shared bottleneck $\mathbf{Z}$ forces downstream predictors to disentangle conflicting task requirements; (2) gradient conflicts on shared parameters lead to the seesaw phenomenon; (3) the architectural transition from attention-based encoding to static MLP scoring prevents end-to-end context-aware ranking and creates computational bottlenecks.

In contrast, OneRank internalizes multi-task reasoning within a unified Transformer architecture through task-specific tokens with mutual invisibility, candidate-aware contextualization, strategic gradient detachment, and dynamic matching-based scoring, enabling superior scaling and stable multi-task optimization.

\section{Conclusion}
\label{sec:conclusion}

In this paper, we identified critical limitations in existing multi-task recommender systems arising from the conventional encoder-predictor separation: task-agnostic information bottlenecks, gradient conflicts leading to the seesaw phenomenon, and architectural transitions that prevent context-aware dynamic ranking and scaling potential. To address these challenges, we proposed OneRank, a unified Transformer-native multi-task ranking framework that internalizes multi-task reasoning within a coherent architectural design. Our approach introduces task-specific tokens with mutual invisibility for early specialization, candidate-aware contextualization via situational descriptors to bridge training-serving gaps, and controlled cross-task attention with strategic gradient detachment for flexible knowledge transfer. By replacing static MLP-based scoring with dynamic matching formulations, OneRank achieves context-aware and task-adaptive ranking within a unified paradigm. Extensive offline experiments and large-scale online A/B testing demonstrate that OneRank significantly outperforms state-of-the-art baselines across multiple tasks while maintaining computational efficiency, validating the effectiveness of our unified Transformer-native design for scalable industrial deployment.

\begin{acks}
This work is supported in part by National Natural Science Foundation of China (No. 62472427 and No. 62422215), Beijing Outstanding Young Scientist Program NO.BJJWZYJH012019100020098, Intelligent Social Governance Platform, Major Innovation \& Planning Interdisciplinary Platform for the ``Double-First Class'' Initiative, Renmin University of China, Public Computing Cloud, Renmin University of China, fund for building world-class universities (disciplines) of Renmin University of China, Intelligent Social Governance Platform. 
\end{acks}

\bibliographystyle{ACM-Reference-Format}
\bibliography{sample-base}

\appendix

\section{Offline Experimental Setup}
\label{appendix:implementation}

\subsection{Dataset Details}

Existing public datasets do not provide the combination of rich sequential item features, explicit multi-task annotations, and ranking-stage candidate sets required for realistic industrial model evaluation. Therefore, we conduct offline experiments on a large-scale proprietary dataset collected from Shopee, a leading e-commerce platform serving billion-scale users across Southeast Asia and Latin America. The dataset is constructed from 30 consecutive days of user interaction logs in December 2025 and is specifically designed for multi-task ranking, covering click, add-to-cart, and order feedback signals.

\subsection{Evaluation Metrics}

We assess model performance under three types of user feedback signals: click (C), add-to-cart (A), and order (O). For each feedback type, following prior work~\cite{chang2023pepnet,zhou2018deep,feng2024long}, we report both AUC and GAUC, which are denoted as C-AUC/C-GAUC, A-AUC/A-GAUC, and O-AUC/O-GAUC, respectively.

\begin{table*}[t]
\centering
\caption{Performance comparison under different loss weight ratios between InfoNCE and BCE in Eq.~\eqref{eq:final_loss}.}
\label{tab:loss_weight}
\begin{tabular}{ccccccc}
\toprule
\textbf{InfoNCE : BCE} & C-AUC & C-GAUC & A-AUC & A-GAUC & O-AUC & O-GAUC \\
\midrule
1 : 2 & 0.7851 & 0.7795 & 0.8384 & 0.7975 & 0.8938 & 0.8292 \\
1 : 1 & \textbf{0.7910} & \textbf{0.7843} & \textbf{0.8463} & \textbf{0.8036} & \textbf{0.9024} & \textbf{0.8350} \\
2 : 1 & 0.7881 & 0.7826 & 0.8440 & 0.8023 & 0.9006 & 0.8350 \\
\bottomrule
\end{tabular}
\end{table*}

\section{Online A/B Testing Details}
\label{appendix:online}

\subsection{Online Deployment Details}

OneRank is deployed within Shopee’s standard multi-stage ranking pipeline and integrated through a score fusion strategy that combines multiple task outputs into a unified ranking score:
\begin{equation}
\label{eq:score_fusion}
s = a \cdot p_{\text{ctr}} \cdot p_{\text{cvr}} \cdot \text{price}
  + b \cdot p_{\text{ctr}} \cdot \text{ecpm}
  + c \cdot \text{relevance},
\end{equation}
where $p_{\text{ctr}}$ and $p_{\text{cvr}}$ denote the predicted click-through rate and conversion rate, respectively. The first term explicitly optimizes gross merchandise value (GMV), the second term accounts for advertising revenue, and the third term enforces search relevance and user intent alignment. Coefficients $a$, $b$, and $c$ are tuned to balance user experience and business objectives in production.

In online inference, each request may involve up to 4,096 candidate items. To achieve a favorable trade-off between ranking quality and computational efficiency, candidates are partitioned into $8 \times 512$ groups, which are scored in parallel. Each group independently performs cross-attention-based ranking, effectively scaling OneRank to large candidate pools while preserving context-aware list modeling.

\subsection{Evaluation Protocol and Metrics}

Online A/B testing is conducted over a 7-day period from January 8 to January 14, 2026. Following standard industrial evaluation practices, we allocate $10\%$ of live traffic to the treatment group using OneRank and $10\%$ to a baseline control group. Following previous work~\cite{dai2025onepiece}, We evaluate online performance along two complementary dimensions:
\begin{itemize}[leftmargin=0.5cm]
    \item \textbf{Platform Benefits}, including GMV/UU (gross merchandise value per user), Paid Orders/UU (average number of completed paid orders per user, excluding refunds), and AR/UU (advertising revenue per user).
    \item \textbf{User Experience}, measured by Bad Query Rate, defined as the proportion of user queries judged as irrelevant, which serves as a proxy for recommendation accuracy and user satisfaction.
\end{itemize}

\begin{figure}[t]
    \centering
    \includegraphics[width=\linewidth]{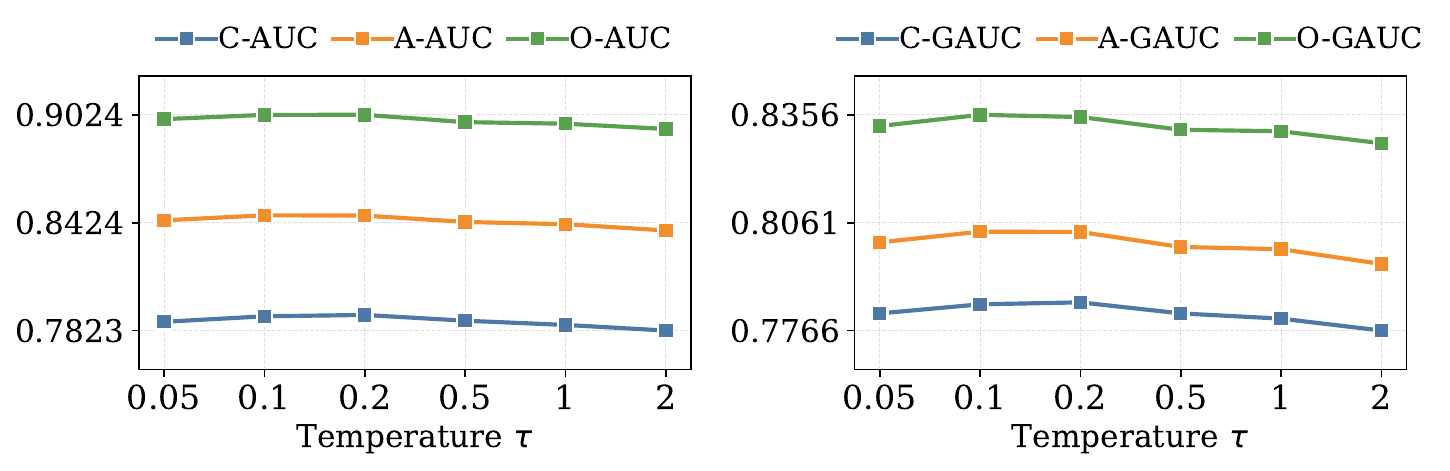}
    \vspace{-20pt}
    \caption{Performance analysis \textit{w.r.t.} temperature.}
    \label{fig:temp_auc_gauc}
    \vspace{-5pt}
\end{figure}

\section{Parameter Sensitivity Analysis}

\paragraph{Performance w.r.t. Temperature in Eq.~\eqref{eq:loss_infonce}}
We analyze the sensitivity of OneRank to the temperature parameter used in the InfoNCE loss (Eq.~\eqref{eq:loss_infonce}). Figure~\ref{fig:temp_auc_gauc} reports the performance trends across different temperature settings. We observe that model performance consistently improves as the temperature decreases from large values, reaches its peak around $0.2$, and slightly degrades when the temperature becomes too small. This trend is consistent across both AUC and GAUC metrics and holds for click, add-to-cart, and order prediction tasks. This behavior aligns with the role of temperature in contrastive learning. A large temperature overly smooths the similarity distribution, weakening discriminative supervision among candidates, while an excessively small temperature sharpens the distribution and may amplify noise or hard negatives, leading to suboptimal optimization. A moderate temperature (around $0.2$) strikes a balance between discrimination strength and training stability, resulting in the best overall performance. Based on this analysis, we set the temperature to $0.2$.

\paragraph{Performance w.r.t. Loss Weight in Eq.~\eqref{eq:final_loss}}

We further study the impact of the relative weighting between the list-wise InfoNCE loss and the point-wise BCE loss in Eq.~\eqref{eq:final_loss}. Table~\ref{tab:loss_weight} reports performance under different loss weight ratios. We observe that assigning equal weights to InfoNCE and BCE consistently yields the best results across all tasks and evaluation metrics. When the BCE loss is over-weighted (1:2), performance degrades, suggesting that relying excessively on point-wise supervision limits the model’s ability to capture relative ranking signals among candidates. Conversely, emphasizing InfoNCE (2:1) improves ranking quality compared with the BCE-dominant setting but still underperforms the balanced configuration. These results indicate that list-wise discrimination and point-wise calibration play complementary roles in OneRank, and a balanced combination of the two objectives leads to the most stable and effective optimization. Based on this analysis, we adopt an equal weighting strategy.

\end{document}